\documentclass[%
 nofootinbib,reprint,
 amsmath,amssymb,
 aps,floatfix,
 longbibliography
]{revtex4-1}

\pdfoutput=1 % if your are submitting a pdflatex (i.e. if you have
\usepackage{graphicx}% Include figure files
\usepackage{dcolumn}% Align table columns on decimal point
\usepackage{bm}% bold math
\usepackage{cancel}
\begin{document}

%\preprint{APS/123-QED}

\title{Einselection, Equilibrium and Cosmology}% Force line breaks with \\

\author{Andreas Albrecht}
 \email{ajalbrecht@ucdavis.edu}
  \author{Rose Baunach}
 \email{baunach@ucdavis.edu}
\affiliation{Center for Quantum Mathematics and Physics and Department of Physics and Astronomy\\ UC Davis, One Shields Ave, Davis CA.}%Lines break automatically or can be forced with \\

\author{Andrew Arrasmith}%
 \email{aarrasmith@lanl.gov}
\affiliation{Theoretical Division, Los Alamos National Laboratory, Los Alamos, NM USA.}


\date{\today}% It is always \today, today,

\begin{abstract}
Our observed Universe has a very strong arrow of time rooted in its low entropy starting point. This low entropy start can be related to various ``tuning puzzles'' about the early state of the Universe.  Here we explore the relationship between the arrow of time and the emergence of classical from quantum in the hopes of ultimately gaining insights into cosmological initial conditions. Our focus is on einselection, the process whereby interactions with an environment select preferred states for a quantum system.  This process plays an essential role in the emergence of classical from quantum. Studies of einselection have so far been limited to cases that exhibit an arrow of time. Here we study the ability of equilibrium systems to exhibit einselection---and investigate whether  detailed balance prevents this---motivated by the question of whether classicality requires an arrow of time.  We present calculations in the adapted Caldeira-Leggett model which demonstrate that einselection can indeed take place in equilibrium systems, and show how this phenomenon is tied to histories which express an arrow of time, despite the global equilibrium.  We discuss some interesting implications of our results for cosmology and cosmological initial conditions. We are intrigued and a bit surprised by the role the consistent histories formalism has ended up playing in our analysis. 
% \begin{description}
% Secondary publications and information retrieval purposes.
% \item[PACS numbers]
% May be entered using the \verb+\pacs{#1}+ command.
% You may use the \texttt{description} environment to structure your abstract;
% use the optional argument of the \verb+\item+ command to give the category of each item. 
% \end{description}
\end{abstract}

\pacs{Valid PACS appear here}% PACS, the Physics and Astronomy
\maketitle

%\tableofcontents

\section{\label{sec:Intro}Introduction}

A quantum system coupled to an environment will generically exhibit entanglement between the system and environment. The onset of such entanglement is called decoherence.  The process of decoherence will cause un-entangled initial states (products in the system-environment partition) to evolve into entangled states, where the system and environment are each described by density matrices (even in the case where the global evolution is unitary and the total state remains pure).  Under certain conditions, which are very common in nature, the density matrix that emerges for the system has eigenstates drawn from a preferred stable set called ``pointer states''.  The process whereby special pointer states are dynamically selected by decoherence is called ``einselection.'' This process plays an essential role in the emergence of classical behavior in quantum systems, for example by rapidly turning ``Schr\"odinger cat'' superpositions into classical mixtures. 

So far einselection has only been studied in the literature (or for that matter in nature) under conditions which exhibit an arrow of time (expressed by the increase of entanglement entropy between the system and environment, for example). This invites the question whether an arrow of time is required for classical behavior to emerge. 

To examine this question further, consider a quantum system in equilibrium, which does not exhibit a global arrow of time by definition. If one considers the detailed balance exhibited by equilibrium systems, it would seem that both entangling and the time reverse (disentangling) would be happening simultaneously, preventing a clear path to einselection from emerging. If such a result was confirmed, it might imply the necessity of an arrow of time to obtain classical behavior. That implication would have interesting consequences for cosmology and various ``tuning puzzles,'' since the arrow of time we experience originates from the low entropy initial conditions of the Universe.  %%%%%%% Changes here for referee response
(Linking the low entropy of the early Universe to special properties of the metric was pioneered by Penrose in the context of his Weyl curvature hypothesis~\cite{Penrose1979}.) 

In this work we have indeed found a link between the emergence of classicality and the arrow of time, although it is not the simple one we anticipated.  Our studies uphold the connection between the arrow of time and einselection, but rather than eliminating the possibility of classical behavior under equilibrium conditions, our explorations of einselection have helped us identify consistent histories which exhibit an arrow of time within the overall equilibrium state.  

Note that it is often routine for physicists to think of equilibrium systems as part of a larger picture (a laboratory for example) in which there is a robust arrow of time. In such situations one can consider processes such as measurements of the system, decoherence etc. which all rely on this arrow of time to operate. In our Universe the origin of this ``laboratory'' arrow of time is cosmological, and it is ultimately the cosmological arrow of time we wish to study here, without any a priori assumption about an external environment with an arrow of time. This motivates our use of the consistent histories formalism, as we discuss below. 

 Our primary tool in this work is the adapted Caldeira-Leggett (ACL) model, which we developed in~\cite{ACLintro} specifically to allow calculations which do not assume an arrow of time from the outset.  This is an important difference from the standard master equation treatments associated with studies of einselection.  As with the original Caldeira-Leggett model, the ACL model describes a simple harmonic oscillator (SHO) coupled to an environment.  We evolve the complete SHO-environment ``world'' fully unitarily, using the highly accurate numerical methods reported in~\cite{ACLintro}.  Our techniques allow us to probe all aspects of the behavior of this model.  Reference~\cite{ACLintro} can also serve as an introduction to the ideas of einselection in the context of the ACL model.  A more general review can be found in~\cite{schlosshauer2007decoherence}.  Both of these resources provide extensive references to the original literature. 
 
 We note that the notion of equilibrium plays something of a dual role in our discussions.  Its main role is as a good example of a physical state which does not exhibit an arrow of time.  In this role it is something of a ``straw man'' which allows us to explore the nature of einselection under conditions not previously studied, and examine the role of time's arrow.  Certainly an equilibrium state is not the only state that can play such a role, and we explicitly expand our discussion to other cases in Appendix~\ref{sec:EEH}.  Separately, one can be curious about the possibility that the universe is globally in a state of equilibrium, and our observed Universe is some sort of fluctuation.   An equilibrium state is certainly expected to be the long-term condition for any finite system (no matter how large), and perhaps other systems as well.  So studies of equilibrium in our toy model might be relevant for assessing such cosmological scenarios.  We will explore these angles later in the paper (where we also acknowledge the range of challenges faced by equilibrium cosmological models). 
 
Quantum physics is a topic which can generate fraught discussions about interpretation.  Most actual calculations are disconnected from those discussions, following standard conventions that produce uncontroversial mathematical results which in most cases are straightforward to connect with data. Consideration of cosmological questions, in which there is no external observer, can sometimes require a more concrete stand on interpretation.  In this paper we ultimately work with the consistent histories formalism, which allows analysis of quantum systems without reference to an external observer (see~\cite{Hartle:2020fis} for some recent reflections on these issues). 

 The consistent histories (CH) formalism has the well-known feature that there are generally many alternate sets of histories available for interpreting the same quantum system.  While the formalism is able to assign relative probabilities to histories \emph{within} a given set, it is agnostic about how one is to make choices among the different sets.  This feature has spawned diverse responses. Some are content to accept this ambiguity as part of the nature of quantum physics, while others seek to add requirements beyond the CH formalism to pare down the possibilities.  Still others feel this feature is grounds for skepticism about the entire CH formalism. We are intrigued by how this intrinsic ambiguity in the CH formalism enables an understanding of the way an equilibrium system can exhibit both detailed balance \emph{and} einselection (along with the associated arrow of time). 

This paper reports substantial technical work using the ACL model. The reader who is mainly interested in the conclusions we draw for cosmology may wish to start by reading Sect.~\ref{sec:cosmo}.  The full structure of this paper is as follows: 
In the next two sections we give a fairly conventional treatment, which will work as a precursor to using the CH formalism.  In Sect.~\ref{sec:eqm} we introduce the basics of the ACL model and explain how we construct the equilibrium state we use throughout the rest of the paper.  In Sect.~\ref{sec:einselect} we explore einselection in our equilibrium system using standard tools based on correlation functions. We conclude Sect.~\ref{sec:einselect} by noting how the standard tools implicitly assume an external environment with an arrow of time. That motivates the extension of our results to a full treatment with the CH formalism which we undertake in Sect.~\ref{sec:CH}.  This expands our understanding of the nature of the einselection and allows us to more fully examine the role of the equilibrium assumption. The CH formalism also allows us to take a closer look at the relationship between einselection and the arrow of time in our calculations.  This we do in Sect.~\ref{sec:AOT}, where we find that the histories we use to study einselection come with a built-in arrow of time. We interpret this arrow
% of time 
in terms of fluctuations of the equilibrium system, and also study its relationship to the well-known feature that the consistent histories formalism usually describes multiple sets of consistent histories which coexist as alternate and disconnected descriptions of the same system. Section~\ref{sec:AOT} is where our concrete technical conclusions for the ACL model are presented. 

To facilitate the application of our results to initial conditions for cosmology, in Sect.~\ref{sec:icch} we offer some general observations about the role of initial conditions in the CH formalism.  
Finally, in Sect.~\ref{sec:cosmo} we relate the insights that have emerged from this work to cosmological questions, especially as they pertain to cosmological initial conditions.  We give some additional attention to the de Sitter Equilibrium cosmological models, but most of our reflections are of a more general nature. We have tried to make Sect.~\ref{sec:cosmo} a self-contained account of our insights and main conceptual points. We outline our main conclusions in Sect.~\ref{sec:conclude}.  Due to disagreements among experts even about what makes a good theory of initial conditions (which we review in Sect.~\ref{sec:CosmoBackground}), our discussions of the implications of our work for cosmology are necessarily more open ended than the concrete technical discussions of the ACL results.   

Appendix~\ref{sec:RanPhase} examines the robustness of our model and our definition of equilibrium. Appendix~\ref{sec:EEH} extends our results to the case where we put the global system in an eigenstate of the total Hamiltonian, suggesting an ``Eigenstate Einselection Hypothesis'' akin to the well-known ``Eigenstate Thermalization Hypothesis.''

\section{\label{sec:eqm}The ACL model and equilibrium}
The ACL model describes a system coupled to an environment. The ACL Hamiltonian is given by
\begin{equation}
H_w = H_s \otimes {\bf{1}}_{}^e + {q_s}\otimes H_e^I + {{\bf{1}}^s} \otimes {H_e}
\label{eqn:Hform}
\end{equation}
where $s$ and $e$ refer to the system and environment.  The system is a SHO, truncated in a particular way to enable stable numerical computation.  The second term in Eqn.~\ref{eqn:Hform} is the interaction term, where $q_s$ is the position operator of the SHO and
\begin{equation}
H_e^I = {E_I}R_I^e + E^0_I.
\label{eqn:Hiedef}
\end{equation}
The self Hamiltonian of the environment is given by
\begin{equation}
H_e = {E_e}R^e + E^0_e.
\label{eqn:Hedef}
\end{equation}
The matrices $R^e$ and $R^e_I$ are independently constructed random Hermitian matrices which are held constant throughout a given calculation\footnote{Each independent matrix element is drawn uniformly from the interval $[-0.5,0.5]$.  We have checked that (basically due to the central limit theorem) this is equivalent to drawing the random numbers from a normal distribution for the values of $N_e$ we consider.}. In~\cite{ACLintro} we provide full details of the ACL model, demonstrate its ability to reproduce standard results from the decoherence and einselection literature, and also demonstrate the ability of the ACL model to evolve into an equilibrium state.  

\begin{figure}
    \centering
    \includegraphics{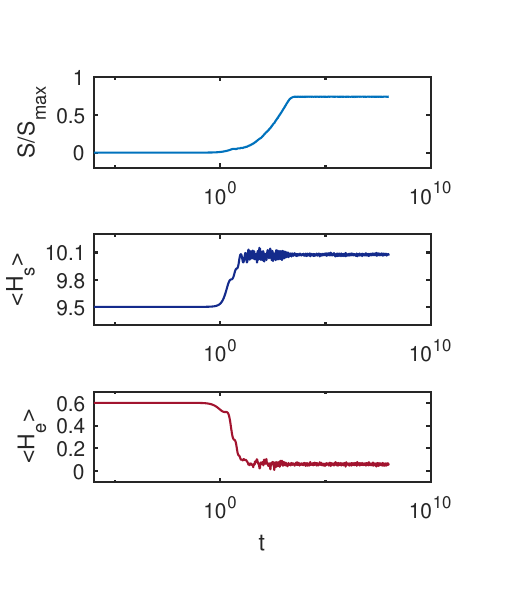}
    \caption{Entropy (top panel), SHO energy (middle) and environment energy (bottom). By $t=3\times 10^6$ these curves have stabilized, supporting the case that the global state at this time represents an equilibrium state.  We use this state in our subsequent calculations. }
    \label{fig:equillibration}
\end{figure}

Figure~\ref{fig:equillibration} shows the evolution of the von Neumann entropy and system and environment energies for a product initial state.
Initially the entropy grows and energy flows between $s$ and $e$, but later equilibration occurs: After $t\approx 10^3$ there is no net flow of energy and the entropy holds steady, up to small fluctuations.  The equilibrium state used throughout this paper is arrived at by tracking this evolution and taking a snapshot of the state of the entire system at $t=3\times 10^6$, well into the equilibrium phase. We write this equilibrium state (in the full $w = s \otimes e$ space) as $\left| \mathcal{ E} \right\rangle $.\footnote{The energy curves in Fig.~\ref{fig:equillibration} show a noisy period as equilibrium fully sets in which gives the appearance, if closely scrutinized, that energy might not be completely conserved. This is an artifact of the energy in the interaction term of $H_w$ not being shown.  Our techniques insure that the full energy of $w$ is conserved to machine precision~\cite{ACLintro}.}  

For Fig.~\ref{fig:equillibration} we use $E^e_I = 0.01$, $E_e =  0.05$, $E^0_I = E^e_I$ and $E^0_e=E^e$.  The initial state is a product of the $\alpha=3$ coherent state for $s$, and the $i=500$ eigenstate of $H_e$ for $e$.  The subsystem dimensions are $n_s=30$ and $n_e=600$. Information about how we approximate an SHO in a finite space, the accuracy of our numerical computations and details of how these states are constructed can be found in~\cite{ACLintro}. Also, in Appendix~\ref{sec:RanPhase} we further scrutinize the notion of equilibrium we use here.

\section{\label{sec:einselect}Einselection}

Einselection is related to the robustness of the system states under interaction with the environment.  Several standard approaches were used in~\cite{ACLintro} to study einselection.  Here we utilize a scheme related to the ``predictability sieve'' approach (a scheme developed by Zurek and collaborators~\cite{Zurek1993predictsieve, Zurek:1992mv} and applied in~\cite{ACLintro}, where we give more extensive reference).  We pose the conditional probability question, ``if the SHO is found in state ${\left| {\psi \left( {{t_0}} \right)} \right\rangle _s}$ at $t_0$, what is the probability of finding the system in ${\left| {\psi \left( {{t_1}} \right)} \right\rangle _s}$ at $t_1$?'' where 
\begin{equation}
    {\left| {\psi \left( {{t_1}} \right)} \right\rangle _s} = \exp(-i(t_1-t_0)H_s/ \hbar){\left| {\psi \left( {{t_0}} \right)} \right\rangle _s}.
    \label{eqn:psi2def}
\end{equation}

To address this question we use the projection operators
\begin{equation}
    {P_0} \equiv {\left| {\psi \left( {{t_0}} \right)} \right\rangle _s}_s\left\langle {\psi \left( {{t_0}} \right)} \right| \otimes {{\bf{1}}_e}
    \label{eqn:p1def}
\end{equation}
and
\begin{equation}
    {P_1} \equiv {\left| {\psi \left( {{t_1}} \right)} \right\rangle _s}_s\left\langle {\psi \left( {{t_1}} \right)} \right| \otimes {{\bf{1}}_e}
    \label{eqn:p2def}
\end{equation}
and construct
\begin{equation}
    \left| \widetilde{{1,0}} \right\rangle  \equiv {P_1}T\left( {{t_1} - {t_0}} \right){P_0}\left| \mathcal{E} \right\rangle  \times {\left( {\left\langle \mathcal{E} \right|{P_0}\left| \mathcal{E} \right\rangle } \right)^{ - 1/2}}
    \label{eqn:12def}
\end{equation}
where $\left| \mathcal{ E} \right\rangle $ is the equilibrium state in the full $w = s \otimes e$ space and 
\begin{equation}
    T\left( {\Delta t} \right) \equiv \exp \left( { - i{H_w}\Delta t / \hbar} \right).
    \label{eqn:Tdef}
\end{equation}
With these definitions, 
\begin{equation}
    {p_{10}}\left( {\Delta t} \right) \equiv \left\langle {\widetilde{1,0}}
 \mathrel{\left | {\vphantom {{1,0} {1,0}}}
 \right. \kern-\nulldelimiterspace}
 {\widetilde{1,0}} \right\rangle 
 \label{eqn:p12def}
\end{equation}
is the quantity which answers the conditional probability question posed. (Note the appearance of a normalization factor in Eqn.~\ref{eqn:12def} which produces the standard normalization used when constructing conditional probabilities.)

\begin{figure}
    \centering
    \includegraphics{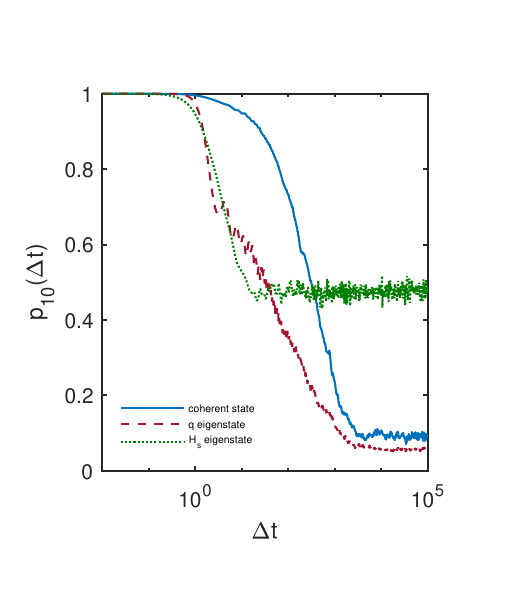}
    \caption{These three curves (defined in Eqn.~\ref{eqn:p12def}) indicate the stability of specific SHO states against interaction with the environment. The coherent state (solid curve) remains stable longer than the position eigenstate (dashed) or the energy eigenstate (dotted), indicating that the coherent state is einselected over the others. (The SHO period is $2\pi$ in these units.) }
    \label{fig:ForwardThree}
\end{figure}

Figure~\ref{fig:ForwardThree} shows $p_{10}$ for ${\left| {\psi \left( {{t_0}} \right)} \right\rangle _s}$ chosen to be either a coherent state (with $\alpha=3$), an eigenstate of $q_s$ (situated at a location similar to the position of the $\alpha=3$ coherent state) or the $n=7$ eigenstate of $H_s$ (which has a similar energy to the other states used here).  
One can see that for a period of time one is certain to find the SHO in the state time evolved from its initial state by $H_s$.  This is the period during which $p_{10}$ stays at unity. Eventually the interactions take their toll, and the state of the SHO has less and less overlap with the state it would have had if it were decoupled from the environment. This phase is manifested by decreasing values of $p_{10}$. The fact that $p_{10}$ remains close to unity for much longer in the coherent state case shows that the coherent states are more stable against decoherence with the environment\footnote{The fact that for the $H_s$ eigenstate case $p_{10}$ levels off at around $0.5$ suggests that at late times our system may be approaching the quantum limit, as discussed in Appendix A of~\cite{ACLintro}. }.  This is a situation often found in nature, which we realize in the ACL model by appropriate choices for the various parameters. (In~\cite{ACLintro} we show how different parameter choices in the ACL model can lead to different pointer states being einselected, but here we stay in the limit where the coherent states are the pointer states.)

\begin{figure}
    \centering
    \includegraphics{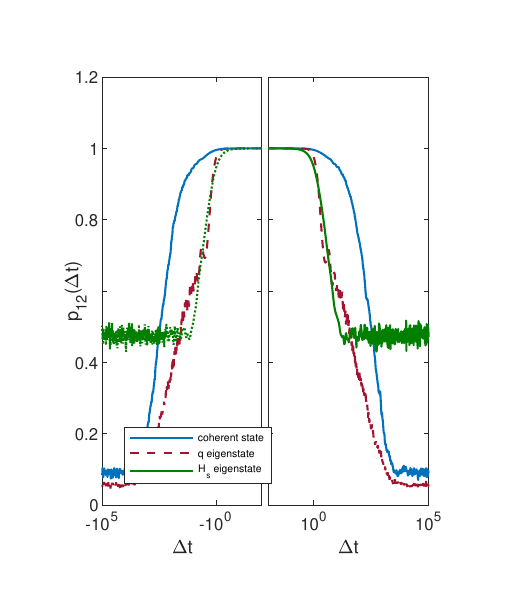}
    \caption{Figure~\ref{fig:ForwardThree} is shown in the right panel, with the same quantities evaluated for negative values of $\Delta t$ in the left panel.  Taken together, these curves reflect the specific states cohering out of equilibrium, becoming fully cohered at $\Delta t = 0$, and then decohering back as $\Delta t$ takes increasing positive values.  The approximate time symmetry that appears here is expected given that the primary condition is placed at $\Delta t = 0$.  These results will contribute to our more thorough discussion of the arrow of time in Sect.~\ref{sec:AOT}.  }
    \label{fig:BothwaysThree}
\end{figure}

One can also evaluate $p_{10}$ for negative values of $\Delta t$.  This corresponds to probing the SHO state at times prior to $t_0$, where the $P_0$ condition is imposed. Figure~\ref{fig:BothwaysThree} 
shows $p_{10}$ for both negative and positive values of $\Delta t$. The results appear to show the chosen SHO state ``cohering'' out of equilibrium into the chosen state at $\Delta t = 0$ and then decohering back toward equilibrium\footnote{Generally, the phenomena which degrade the correlations include both decoherence and dissipation.  In everyday macroscopic systems decoherence operates on a much faster timescale and is the focus of discussions of stability and einselection.  We've shown in~\cite{ACLintro} that both phenomena are present in the ACL model, although the timescales are much closer together (as one might expect in a finite toy model).}.  We will come back to this picture when studying the system from the point of view of consistent histories. 

Naively, it would appear that we have demonstrated that einselection can indeed happen in equilibrium systems, thus answering in the affirmative a question which motivated this paper. However, the formalism we developed in this section requires more scrutiny. A standard interpretation would say that the projection operators we use here would describe measurements of the SHO by some apparatus external to both $s$ and $e$.  Such a measurement could be expected to throw the whole thing out of equilibrium, so it is not clear if we have really addressed the original question.  

Specifically, after operating with $P_0$, one is left with a product state with zero entanglement entropy between the system and environment. The only aspect that reflects the fact that we started with the equilibrium state $\left| \mathcal{ E} \right\rangle $ is the specific environment state which multiplies the system state ${\left| {\psi \left( {{t_0}} \right)} \right\rangle _s}$ (determined by $P_0$).   The subsequent evolution is that of an initial product state such as shown in Fig.~\ref{fig:equillibration} and studied extensively in~\cite{ACLintro}. Motivated by these considerations, we now turn to the consistent histories formalism which allows an analysis which looks less like an external disruption of our equilibrium system.  We will see how this formalism introduces some new considerations to our assessment of einselection. 

\section{\label{sec:CH}Consistent Histories}

The consistent histories (CH) formalism is a tool for identifying classical behavior in a closed quantum system without reference to an outside observer. It was proposed in 1984~\cite{Griffiths:1984rx}, and since then a substantial literature has emerged (see for example~\cite{Gell-Mann:2018dzd,Albrecht:1992rs,Albrecht:1992uc,Paz:1993tg,PhysRevD.47.3345,Dowker:1994ac,omnes1999understanding,Hartle:2006iq,Hartle:2008mv,Halliwell:2017anw,Halliwell:2003fw,Hartle:2020his} and for a recent review see~\cite{RevModPhys.82.2835}).  We use the formalism here in a very similar manner to the way it is used in~\cite{Albrecht:1992rs}.  The next subsection sets up our techniques in a way that might serve as a very brief introduction to the CH formalism, at least in the form we use here.  The subsequent {\em Results} subsection presents results which address the topics of interest in this paper using the CH formalism. This subsection also offers intuitive interpretations of the CH quantities, which may be all some readers need to know about the CH formalism. Such readers might try skipping straight to Sect.~\ref{sec:CHres}.

\subsection{\label{sec:CHdefs} Formalism}
The CH formalism expresses the full time evolution of a quantum system in terms of histories formed using complete sets of projection operators.  We start our discussion by using $P_0$ and $P_1$ from Eqns.~\ref{eqn:p1def} and~\ref{eqn:p2def} to define the complementary projectors 
\begin{equation}
    \begin{array}{l}
{P_{\bcancel{0}}} \equiv {\bf{1}} - {P_0}\\
{P_{\bcancel{1}}} \equiv {\bf{1}} - {P_1}.
    \end{array}
    \label{eqn:pbardef}
\end{equation}
We consider the time evolution given by 
\begin{eqnarray}
\left| {\psi \left( {{t_1}} \right)} \right\rangle 
 &=& T\left( {{t_1} - {t_0}} \right)\left| {\psi \left( {{t_0}} \right)} \right\rangle \\
  &= &{\bf{1}}T\left( {{t_1} - {t_0}} \right)  {\bf{1}}\left| {\psi \left( {{t_0}} \right)} \right\rangle.
 \end{eqnarray}
 with $T$ defined in Eqn.~\ref{eqn:Tdef}. (Here the states and operators are in the full $w=s\otimes e$ space.) Since ${P_1} + {P_{\bcancel{1}}} = {P_0} + {P_{\bcancel{0}}} = {\bf{1}}$ (thus forming ``complete sets''), one can continue by writing
 \begin{eqnarray}
 \left| {\psi \left( {{t_1}} \right)} \right\rangle &= &\left( {{P_1} + {P_{\bcancel{1}}}} \right)T\left( {{t_1} - {t_0}} \right) \nonumber \\
 & \times & \left( {{P_0} + {P_{\bcancel{0}}}} \right)\left| {\psi \left( {{t_0}} \right)} \right\rangle \\
 &=&  {P_1}T\left( {t_1} - {t_0}\right){P_0} \left| {\psi \left( {{t_0}} \right)} \right\rangle \nonumber \\
 & + & {P_1}T\left( {t_1} - {t_0}\right){P_{\bcancel{0}}}\left| {\psi \left( {{t_0}} \right)} \right\rangle \nonumber \\
  & + &   {P_{\bcancel{1}}}T\left( {t_1} - {t_0}\right){P_0}\left| {\psi \left( {{t_0}} \right)} \right\rangle \nonumber \\
 & + & {P_{\bcancel{1}}}T\left( {t_1} - {t_0}\right){P_{\bcancel{0}}}\left| {\psi \left( {{t_0}} \right)} \right\rangle \label{eqn:allprojsin} \\
 &\equiv & \left| {1,0} \right\rangle  + \left| {1,\bcancel{0}} \right\rangle  + \left| {\bcancel{1},{0}} \right\rangle  + \left| {\bcancel 1,\bcancel 0} \right\rangle 
 \label{eqn:ppsdef}
\end{eqnarray}
where the quantities in Eqn.~\ref{eqn:ppsdef} are defined by
\begin{equation}
\left| {i,j} \right\rangle  \equiv {P_i}T\left( {{t_1} - {t_0}} \right){P_j}\left| {\psi \left( {{t_0}} \right)} \right\rangle.
\label{eqn:ppsdefExpanded}
\end{equation}
Note that Eqns.~\ref{eqn:12def} and~\ref{eqn:ppsdefExpanded} are related by
\begin{equation}
    \left| \widetilde{i,j} \right\rangle = \left| {i,j} \right\rangle {\left( {\left\langle  {\psi \left( {{t_0}} \right)} \right|{P_0}\left| {\psi \left( {{t_0}} \right)}  \right\rangle } \right)^{ -1/2}},
    \label{eqn:tildeProp2notilde}
\end{equation}
meaning that these two quantities just differ by a normalization. 

Equation~\ref{eqn:ppsdef} amounts to organizing the time evolution in terms of paths or histories, where each term in Eqn.~\ref{eqn:ppsdef} represents a different history determined by which projections are chosen at each of the two times\footnote{This construction has the look of a derivation of the path integral, but in the consistent histories formalism there is generally no expectation that the usual continuum limits need be taken.}. In general the consistent histories formalism can accommodate any number of times where complete sets of projection operators are inserted, as well as more finely grained sets of projections themselves. Here we stick to using only two projection times ($t_0$ and $t_1$), and a very simple choice of projectors ($P_1$, $P_0$ and their compliments).  These will suffice to explore the physical questions of interest while keeping our formalism and computations as simple as possible. 

Next we define the ``Decoherence functional''
\begin{equation}
    {D_{ij,kl}} \equiv \left\langle {{i,j}}
 \mathrel{\left | {\vphantom {{i,j} {k,l}}}
 \right.}
 {{k,l}} \right\rangle
 \label{eqn:Ddef}
 \end{equation}
 (with $\left| {i,j} \right\rangle $ defined in Eqn.~\ref{eqn:ppsdefExpanded})\footnote{Technically our $D$ is a  {\em function} of discrete variables, not a function{\em al}, but we stick to the standard usage to avoid generating arcane nomenclature.}.
 The CH formalism seeks to use the diagonal elements of $D$ to assign the probability
\begin{equation}
    p^{CH}_{ij}=D_{ij,ij}
    \label{eqn:pchdef}
\end{equation}
to the $ij$  path.  The paths are considered consistent if the $p^{CH}_{ij}$ obey the sum rules expected of classical probabilities.  For example, one could define a coarse grained history (labeled by $ 1 \circ $) where no projections are inserted at $t_1$, and classically one would expect
\begin{equation}
    ^cp^{CH}_{1 \circ} = p^{CH}_{1 0}+ p^{CH}_{1 {\bcancel{0}}}.
    \label{eqn:pch2circclass}
\end{equation}
However, in general off diagonal elements of $D$ come in giving
\begin{eqnarray}
          ^qp^{CH}_{1 \circ} & = &  ^cp^{CH}_{1 \circ} + D_{10,1\bcancel{0}} + D_{1{\bcancel{0}},1 {0}} .
        \label{eqn:pch2circquant} 
\end{eqnarray}
We say we have identified good sets of consistent histories when the off diagonal contributions in Eqn.~\ref{eqn:pch2circquant} (which describe quantum interference effects) are sufficiently small, so that the classical expression (Eqn.~\ref{eqn:pch2circclass}) is obeyed to the desired tolerance\footnote{We note that, in addition to its use in the CH formalism, Eqn.~\ref{eqn:pch2circclass} shows up in other contexts as a metric of classicality.  For example, Eqn.~\ref{eqn:pch2circclass} corresponds to a form of the Kolmogorov consistency condition, and there is a body of work investigating deviations from Eqn.~\ref{eqn:pch2circclass} in quantum systems and their implications for classicality within a projective measurement framework  (e.g.~\cite{MilzPhysRevX.10.041049,Milz2020,Strasberg2019}). }.   

\begin{figure}
    \centering
    \includegraphics{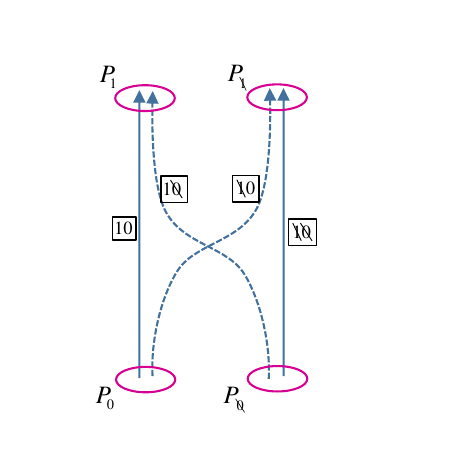}
    \vspace{-0.4cm}
    \caption{A schematic illustrating the four paths constructed in Sect.~\ref{sec:CHdefs}. The projection operators correspond to circles and the path labels are marked in boxes.  In Sect.~\ref{sec:CHres} we focus mainly on the two paths which end at $P_1$ (Fig.~\ref{fig:df245}). One path ($10$, solid) arrives from $P_0$ (giving the simple behavior of a decoupled SHO), the other (${1\bcancel{0}}$, dashed) arrives from $P_{\bcancel{0}}$.  (The ${1\bcancel{0}}$ path would be impossible without interactions with the environment.) }
    \label{fig:Branching}
\end{figure}

Figure~\ref{fig:Branching} illustrates the full set of paths considered here, along with the path labels and corresponding projection operators.
In a more general CH formalism, with many projection times and many components to the complete sets of projectors, there are a multitude of sum rules that can be checked. For our purposes the relatively simple framework set up here suffices. 

The formalism described here is perfectly well formulated for either $t_1 > t_0$ or $t_1 < t_0$.  The subscript refers to the order in which the projections appear in Eqn.~\ref{eqn:allprojsin}, but $T(t_1-t_0)$ is well defined for both positive and negative arguments. Thus the arrows in Fig.~\ref{fig:Branching} really refer to the order of the projections, and one can consider cases where time flows from top to bottom in this diagram.  One can think of the projection at $t_0$ giving initial conditions for the path when $t_1 > t_0$ and as giving final conditions when $t_1 < t_0$\footnote{Use of final conditions in the CH formalism has been discussed for example in~\cite{Gell-Mann:2018dzd,PhysRevD.47.3345}.}. This aspect will be important to the discussion in Sect.~\ref{sec:AOT}.  

\vspace{-0.2cm}

\subsection{\label{sec:CHres}Results}
\begin{figure}
    \centering
    \includegraphics{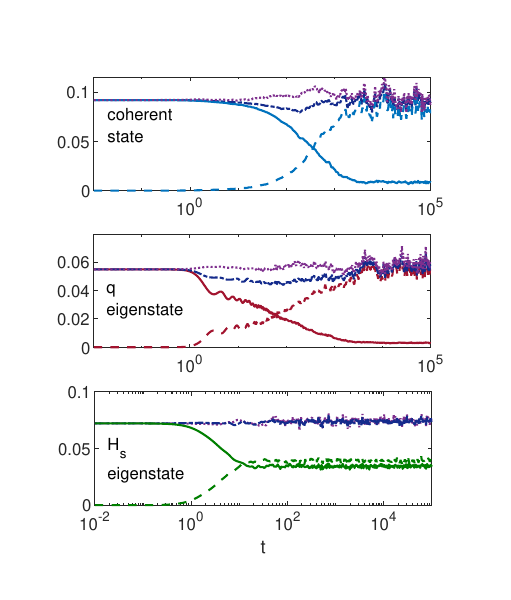}
    \vspace{-0.4cm}
    \caption{A CH treatment of different initial system states as they interact with the environment. The solid curves are $p^{CH}_{10}\left(t\right)$ (which are none other than the correlation functions shown in Fig.~\ref{fig:ForwardThree}). The dashed curves are the same quantity for the alternate $1 {\bcancel{0}}$ path. The two top curves give $^cp^{CH}_{1\circ}\left(t\right)$ and $^qp^{CH}_{1\circ}\left(t\right)$.  The extent to which the top two curves are different from one another signals quantum interference effects between the ${10}$ and $1 {\bcancel{0}}$ paths which undermine attempts to assign classical probabilities.  As discussed in the text, the interference effects do not change our conclusions about einselection for these cases. (A pictorial representation of the paths considered is shown in Fig.~\ref{fig:Branching}.) }
    \label{fig:df245}
\end{figure}

Here we revisit the physical question posed in Sect.~\ref{sec:einselect}---does einselection happen in equilibrium---this time using the CH formalism. We condition on the case where the SHO is found in state $\left | \psi(t_0) \right >_s$ at $t_0$, and compute the probability of finding it in the corresponding evolved state $\left | \psi(t_1) \right >_s$ at $t_1$. While the framework of Sect.~\ref{sec:einselect} implies the measurement of the SHO by an external apparatus, the CH formalism uses projectors to identify paths.  The solid curves in each panel of Fig.~\ref{fig:df245} are called $p^{CH}_{10}$ in the CH formalism, but they are none other than the $p(\Delta t)$ curves shown in Fig.~\ref{fig:ForwardThree}, rescaled according to Eqn.~\ref{eqn:tildeProp2notilde}. Crucially, the CH formalism requires us to consider additional quantities in order to interpret these curves. 
The dashed curve in each panel shows the probability that the SHO was {\em not} in $\left | \psi(t_0) \right >_s$ at $t_0$, but is none the less found in $\left | \psi(t_1) \right >_s$ at $t_1$.  This is the quantity called $p^{CH}_{1{\bcancel{0}}}$ in Sect.~\ref{sec:CHdefs}. The presence of this alternate pathway to $\left | \psi(t_1) \right >_s$ is part of what makes the CH formalism different from our treatment in Sec.~\ref{sec:einselect}. 
Figure~\ref{fig:Branching} illustrates the full set of paths considered here, along with the path labels and corresponding projection operators (Fig.~\ref{fig:df245} only shows information about the two paths which arrive in the upper left of Fig.~\ref{fig:Branching}).

The dot-dashed curve in each panel of Fig.~\ref{fig:df245} is just the sum of the solid and dashed curves ($^cp^{CH}_{1\circ}$ from Eqn.~\ref{eqn:pch2circclass} giving the expected total classical probability), and the dotted curve shows $^qp^{CH}_{1\circ}$ from Eqn.~\ref{eqn:pch2circquant}, which includes quantum interference effects. The degree to which the dot-dashed curves (classical) and the dotted curves (quantum) differ indicates the breakdown of the classical rules for probabilities.  

In Sect.~\ref{sec:einselect} we examined the (rescaled) solid curves from Fig.~\ref{fig:df245} which we presented in Fig.~\ref{fig:ForwardThree}. We used the deviation from constant behavior as a signal of instability under interaction with the environment.  The fact that the coherent state case stayed constant for longer than the other cases led us to conclude that the coherent states were more stable under interactions with the environment, and thus were einselected by these interactions.  Since these same curves appear in the CH discussion, it seems we would draw the identical conclusions using the identical information.  

The new feature that is added by the CH formalism is the chance to check for interference effects among different paths, which can undermine the assignment of classical probabilities to the paths. This sort of breakdown is a physically different way the interactions with the environment can erode classical behavior, and this erosion is signalled by  deviations between the dotted and dot-dashed curves in Fig.~\ref{fig:df245}.  Since the deviations between these two curves appear (to the extent that they occur) around the same time as the solid curves start to deviate from constant values, we can argue that the emergence of interference effects does not change our conclusions about einselection for these particular calculations. (In Appendix~\ref{sec:EEH} we present examples where interference effects do change our conclusions about einselection.)

To make such an argument more carefully, one would need a measure of how large the interference effects need to be to register a breakdown of classicality. If our tolerance was very tight, we might need to zoom in to the early-time parts of the curves in Fig.~\ref{fig:df245} to check for small deviations, and it is possible that these small deviations would not appear in the same time order across the three panels.  If that were the case, our argument about einselection could be undermined. On the other hand, a more lax tolerance of interference effects could regard all the interference effects shown in Fig.~\ref{fig:df245} as inconsequential.  Under those conditions our discussion of einselection would revert completely back to the form it took in Sect.~\ref{sec:einselect}.  

As usual in physics, the choice of which tolerances to use should be grounded in practical considerations related to what we intend to do with the SHO.  For example, if the SHO is intended to represent the pendulum of a clock, the accuracy of the clock would dictate the degree of classicality needed for the pendulum.  Such considerations lie far outside scope of our little toy model. The value we see in our analysis of the ACL model is that it has given us a sufficiently concrete framework for calculations to compel us to carefully organize our ways of thinking about the relationship between einselection and equilibrium. We now turn to a discussion of how what we have reported so far relates to the arrow of time. 

% Each panel shows $p^{CH}_{21}$ (solid), $p^{CH}_{2{\bcancel 1}}$ (dashed), $(p^{CH}_{21} +p^{CH}_{2{\bcancel 1}})$ (dot-dashed) and $p^{CH}_{2\circ}$ (dotted), and has a text label indicating which form of the projectors was used.  The solid curves are the same ones plotted in~\ref{fig:ForwardThree} (rescaled according to Eqn.~\ref{eqn:tildeProp2notilde}), and the longer time stability for the coherent state case points to the same conclusion about einselection drawn from Fig.~\ref{fig:ForwardThree}. 

\section{\label{sec:AOT}Arrow of Time}

\subsection{\label{sec:AOTandE} Time's arrow and einselection}

The results we have presented so far clearly have an arrow of time. By construction, the  $p^{CH}_{10}$'s (solid curves in Fig.~\ref{fig:df245}) correspond to paths in which the SHO and the environment are not entangled at $\Delta t = 0$.  The eventual deviations of $p^{CH}_{10}$ from constant behavior correspond to a ``branching'' as the probability for the ``$10$'' path declines, and the probability for arriving at  $\left | \psi(t_1) \right >_s$ from a state different from  $\left | \psi(t_0) \right >_s$ (given by $p^{CH}_{1 \bcancel{0}}$) increases.  

\begin{figure}
    \centering
    \includegraphics{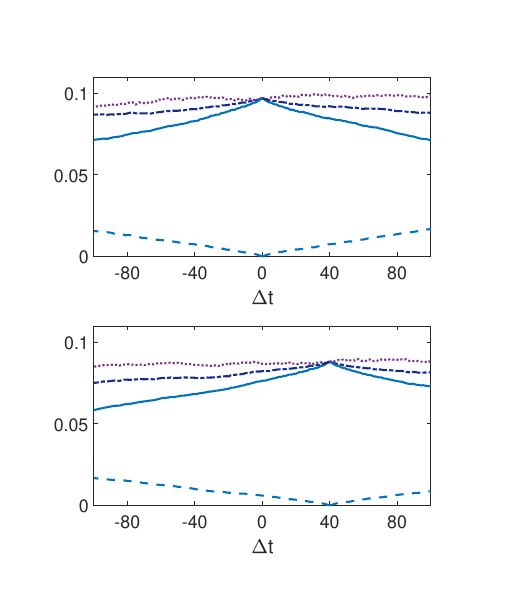}
    \vspace{-0.5cm}
    \caption{Alternate histories: We show the CH quantities from Fig.~\ref{fig:df245} (coherent state case only), but with a linear x-axis and negative values of $\Delta t$ included.  As discussed in the text, each panel shows an alternate CH narrative for the identical quantum state.  The different narratives give conflicting accounts of the arrow of time, both of which are equally valid.}
    \label{fig:2CH}
\end{figure}

As we did with the correlation functions discussed in Sect.~\ref{sec:einselect}, one can consider the $t_1 < t_0$ case, where $P_0$ and $P_{
\bcancel{0}}$ impose \emph{final} conditions. In that case, our formalism explores different histories by which the system can arrive at $\left | \psi(t_0) \right >_s$ from the past.  
Figure~\ref{fig:2CH} shows the quantities given in the top panel of Fig.~\ref{fig:df245}, along with the same quantities evaluated for negative values of $\Delta t$.
Here the x-axis has a linear scale, allowing both negative and positive values of $\Delta t$ to be shown together.   In linking these two paths (positive and negative values of $\Delta t$) together, one is effectively imposing a ``middle condition'' at $t_0$ rather than an initial or final one. 

The top panel of Fig.~\ref{fig:2CH} describes the SHO starting with significant entanglement with the environment at negative $\Delta t$ values.  As $\Delta t$ approaches zero from below, the entanglement approaches zero, and the $10$ and $1{\bcancel{0}}$ paths join together (as $p^{CH}_{1{\bcancel{0}}} \rightarrow 0$). After reaching zero at $\Delta t = 0$, the entanglement increases again and the two paths branch out\footnote{Technically the CH formalism is able to consider branching that is not connected to entanglement with an environment, but such a connection {\em is} present in the cases we consider here.}. The top panel corresponds to the top panel of Fig.~\ref{fig:df245}.  For the bottom panel we've used $t^{'}_0 = t_0+40$ in constructing the projection operators (but still show $\Delta t$ on the x-axis rather than $\Delta t'$, to make our narrative simpler).  What we have done in this case is evolve the equilibrium state to 
\begin{equation}
    \left | \mathcal E \right >^{'} = T(40) \left | \mathcal E \right >
\end{equation}
and used $\left | \mathcal E \right >^{'} $ in the expressions for the CH quantities shown in the lower panel.  Thus, the two panels represent different CH narratives for the identical quantum state. In one case the SHO is in a pure state at $\Delta t = 0$, and becomes more entangled as $ \Delta t$ deviates from zero in either direction.  In this case, the SHO has become significantly entangled with the environment by $\Delta t = 40$.  In the 2nd case, at $\Delta t = 0$ the SHO is in the process of becoming disentangled from the environment, a process which completes at $\Delta t = 40$ and then starts reversing.  Each panel represents a double headed arrow of time, but the time at which the arrow changes direction is different in the two cases. 

Here we have encountered a well-known feature of the CH formalism, namely that there are typically many different sets of histories that coexist as alternate accounts of classical behavior for a given quantum system~\cite{Albrecht:1992uc,Albrecht:1992rs,Paz:1993tg,Dowker:1994dd,Dowker:1994ac}.  The CH formalism on its own is unable assign a preference to one of these sets over another (or assign relative probabilities between the \emph{sets}, even as it does assign relative probabilities to histories drawn from the same set).  As illustrated in Fig.~\ref{fig:2CH}, this ambiguity shows up in the lack of preference for the point in time when the entropy is at a minimum (and its arrow switches directions).

Figure~\ref{fig:2CH} also allows us to revisit the question of detailed balance we raised in the introduction.  There we asked whether the detailed balance properties of equilibrium systems mean that entangling and disentangling processes are happening simultaneously, which would suggest there is no clear route to einselection.  We see that the CH formalism allows us to interpret an equilibrium state with paths which  have separate periods dominated by either entangling or disentangling. On such paths these two processes are \emph{not} happening simultaneously (at least not on an equal basis).  Looking at $\Delta t = 20$ in Fig.~\ref{fig:2CH}, indeed both entanglement and disentanglement are happening ``simultaneously'' in the sense that both processes are represented.  But they are represented on different paths, each of which has a clear direction, and is interpreted as a separate classical description of the behavior of the system. Such paths single out a special time which marks the transition between these two periods, and one might wonder how an equilibrium system can ``choose'' what time that would be.  The answer is that the system does not choose that time, but rather multiple interpretations coexist where the transition between entangling and disentangling occurs at different times.  The multiplicity of the sets of paths (along with the double-headed nature of the arrows) captures the notion of detailed balance, even as the individual paths appear to disregard that notion. 

\subsection{\label{sec:AOTother} Related considerations}

It is standard practice to quantify properties of equilibrium using correlation functions, often time averaged. Indeed this is how we presented earlier versions of this work, for example in~\cite{Albrecht.IPMU.2019}. However, the time averaging and other specifics of those analyses seemed to obscure the relationship between our calculations and traditional ideas about einselection.  We feel the approach we use here offers greater clarity. For one, we see that equilibrium systems can admit descriptions which \emph{do} exhibit an arrow of time.  We find it intriguing that rather than equilibrium conditions preventing the system from exhibiting einselection (as we initially suspected might be the case), exploring the physics of einselection led us to histories which exhibit an arrow of time, even under equilibrium conditions.  While our picture might be described as ``capturing a transient downward, and then upward, fluctuation of the entropy'' in an equilibrium system (certainly a notion commonly associated with a double headed arrow of time), the CH formalism gives a technical account of what such a statement might mean. In particular, it does \emph{not} mean waiting for a recurrence which would bring the full entanglement entropy to a small value.  Rather, it means choosing histories which reflect such a fluctuation.  Such histories which place the fluctuation at any chosen moment in time are equally available.  There is no need to \emph{wait} for any fluctuation, let alone a recurrence.   

In fact, it is straightforward to extend the formalism we've developed here so that the entire set of projectors has the form of Eqn.~\ref{eqn:p1def}.  To do that, one would replace the projectors $P_{\bcancel{i}}$ (defined in Eqn.~\ref{eqn:pbardef}) with sets of $N_s-1$ projectors of the form of Eqn.~\ref{eqn:p1def} using a set of  $\left | \psi_j(t_0) \right >_s$, where $j$ labels a set of states which, along with the original $\left | \psi(t_0) \right >_s$, form an orthonormal basis for $s$.  Creating histories from such projectors would ensure that \emph{every} path had zero entanglement entropy at $t_0$, and that the zero entropy time could be chosen arbitrarily using the ideas discussed in Sect.~\ref{sec:AOTandE}.   The overall (large) entanglement between system and environment would be expressed by nonzero probabilities assigned to many paths, but the entropy on each would be zero at that moment. Generally, the equilibrium nature of the whole system would also show up in the branching behavior we've demonstrated here, which  makes the zero entropy feature only a transient property of the paths which emerge and then decohere according to a double-headed arrow of time. 

While we are on the topic of alternate sets of histories, we should note that the process of einselection itself has long been regarded as a useful tool for selecting a preferred set among the many possible sets of consistent histories~\cite{Albrecht:1992rs,Albrecht:1992uc,Paz:1993tg}.  If the CH projections are made on the pointer states, their robustness leads to greater stability and thus a longer period of classicality.  This is a more formal way of stating the importance of einselection, which we sketched in a more heuristic way in the introduction.  ``Quantum Darwinism''~\cite{2003quant.ph..8163Z, BlumeKohout:2008zz,PhysRevA.93.032126,Arrasmith:2019thu} is another idea for selecting preferred sets of histories. While our toy model is far too simple to illustrate this idea directly, we do not expect that quantum Darwinism could select a preferred set among the histories showing fluctuations at different moments in time, such as those shown in Fig.~\ref{fig:2CH}. 

\begin{figure}
    \centering
    \includegraphics{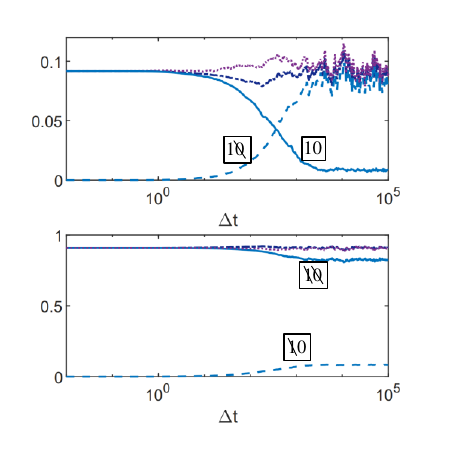}
    \caption{The complete set of histories:  The top panel is the same as the top panel of Fig.~\ref{fig:df245}, and the bottom panel gives the same information for the remaining histories from Fig.~\ref{fig:Branching} (labeled as in Fig.\ref{fig:Branching}).  }
    \label{fig:CHall}
\end{figure}

To further complete our discussion, we present Fig.~\ref{fig:CHall}, which shows the CH quantities for all four paths shown in Fig.~\ref{fig:Branching}.
\begin{figure}
    \centering
    \includegraphics{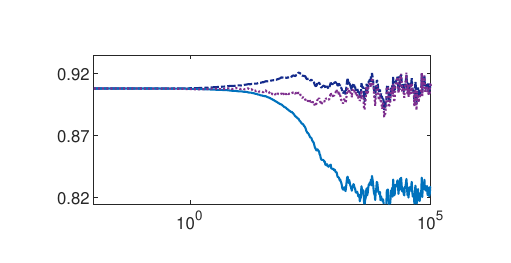}
    \caption{Zooming in on the top curves from the lower panel of Fig.~\ref{fig:CHall}.  Here $y$-axis is on the same scale as the upper panel of Fig~\ref{fig:CHall}, for easier comparison.  One can see that the breakdown of the classical sum rules (indicated by the deviation of the dot-dashed and dotted curves) is of a similar size in both cases.}
    \label{fig:CHallzoom} 
\end{figure}
The additional paths (both of which start with $P_{\bcancel{1}}$) shown in the lower panel exhibit broadly the same features discussed so far for the paths with start with $P_{{1}}$, and reflect the same phenomena.  The main difference is that the probability remains high for the $\bcancel{1}\bcancel{0}$ path, which is not surprising since each of the $P_{\bcancel{i}}$ projectors cover most of the $s$ Hilbert space, and the whole system is in equilibrium (and thus quite spread out in the Hilbert space).  Figure~\ref{fig:CHallzoom} provides a zoomed in picture of part of the lower panel of Fig~\ref{fig:CHall}. This allows us to see that the breakdown of the classical sum rules is happening on similar overall scales for both pairs of paths. 

 Thus far we have presented a variety of results from our toy model related to einselection and the arrow of time in equilibrium systems. We want to explore the implications of these results for cosmological theories, especially theories of cosmic initial conditions. That is the fundamental motivation for this project. Our first step in this direction is to look more generally at the role of initial conditions in the CH formalism. 
 
 \section{\label{sec:icch}The role of initial conditions in the CH formalism}
 We first approached einselection intuitively using a correlation function approach.  We then expanded that discussion using the CH formalism to avoid inadvertently evoking an external observer.   We want to apply the insights we have gained about the presence of einselection in equilibrium, and what that implies for a link between an arrow of time and classical behavior, to the question of cosmological initial conditions.  To facilitate that focus, we first step back and take a broader look at the role of initial conditions in the CH formalism. 
 
 \subsection{\label{sec:mh}Microstate Histories}
 We start by looking at a very special case which we call ``Microstate Histories'' (MH).  It is well known that it is always easy to create consistent histories by choosing projection operators which, unlike those we defined above, project onto microstates of the entire Hilbert space $w$. We write these as
 \begin{equation}
{P_i}^{MH}\left( {{t_0}} \right) \equiv {\left| {{\psi _i}\left( {{t_0}} \right)} \right\rangle _w}_w\left\langle {{\psi _i}\left( {{t_0}} \right)} \right|.
\label{eqn:PMH0def}
 \end{equation}
Furthermore, in the MH scheme projectors at later times, ${P_i}^{MH}\left( {{t_j}} \right)$, are  constructed by using 
 \begin{equation}
  {\left| {{\psi _i}\left( {{t_j}} \right)} \right\rangle _w} \equiv      T\left(t_j-t_0\right){\left| {{\psi _i}\left( {{t_o}} \right)} \right\rangle _w}
  \label{eqn:PMHtdef}  
 \end{equation}
in Eqn.~\ref{eqn:PMH0def}.  One can construct a flawlessly consistent set of histories by letting $i$ run across a complete basis for $w$.  Because $T$ is unitary, the orthonormality of the ${\left| {{\psi _i}\left( {{t_0}} \right)} \right\rangle _w}$ states guarantees that all the off-diagonal elements of the decoherence functional will remain exactly zero in this scheme.  For this scheme to work, we needed to select  $ \left| {{\psi _i}\left( {{t_0}} \right)} \right\rangle_w $
as a microstate in the full Hilbert space of $w$ so that the unitarity of $T$ can be exploited.  If one uses the MH scheme with initial state ${\left| \mathcal I \right\rangle _w}$, the only role for ${\left| \mathcal I \right\rangle _w}$ is to assign the probability 
\begin{equation}
p_i^{MH} \equiv \left|  _w\left\langle {{\psi _i}\left( {{t_0}} \right)} \right.     {\left| \mathcal I \right\rangle _w} \right|^2
\label{eqn:piMH}
\end{equation}
 to the history constructed with projectors $P_i^{MH}$. In the MH scheme there is no guarantee that there would be anything particularly classical about the paths, other than that they (in a trivial way) could be assigned definite probabilities which obey all the classical sum rules.  All the MH scheme does is exploit the fact that ${\left| \mathcal I \right\rangle _w}$  can be expanded in a complete basis, and that if one evolves that basis with $T$, the expansion coefficients will not change over time. 

If the $w$ space had an internal structure complex enough to describe observers and observables (certainly more complicated than our simple toy model), observers on the $i^{\rm th}$ MH path would naturally regard 
${\left| {{\psi _i}\left( {{t_0}} \right)} \right\rangle _w}$ as their initial state, not ${\left| \mathcal I \right\rangle _w}$\footnote{Note that in order to account for the existence of observers and observables one would need to drill down to more fine-grained sets of histories describing these aspects.}.  If any of these observers were cosmologists, they might debate the relative merits of the global state ${\left| \mathcal I \right\rangle _w}$, or perhaps a competing theory ${\left| \mathcal I' \right\rangle _w}$, and the different values of $p^{MH}_i$ they provide.  But aside from certain cosmological considerations, the state ${\left| {{\psi _i}\left( {{t_0}} \right)} \right\rangle _w}$ is all an observer would need to account for the physics they experience on their particular history. 

We have presented the highly idealized MH case to illustrate how in the CH formalism the notion of a global initial state can be quite disconnected from the experiences of  observers on a particular history.  We will now turn to a less idealized case and note that even there a similar disconnect is possible.  

\subsection{\label{sec:generalcase}General case}
Typically the projectors used in the CH formalism are not formed from microstates of the entire Hilbert space, but focus on subspaces (as we've done with our toy model), or perhaps use other forms of ``coarse graining.''  This allows one to focus on observables or other quantities of interest while ignoring unobservable microscopic degrees of freedom. Generally, such a focus has a key role in identifying classical behavior. As we've illustrated with the calculations in this paper (and unlike the idealized MH case), for such histories quantum interference among the paths becomes a real issue which must be quantified.  Only histories with sufficiently low interference effects may be assigned classical probabilities.   Still, when the conditions are such that the interference effects are low, then each classical history has a ``life of its own,'' and would naturally identify its initial state with the first projector of that history.  

Careful scrutiny would reveal that in this more general case the disconnect from the global initial state ${\left| \mathcal I \right\rangle _w}$ is not as trivial as in the MH case, but in the end it could appear to take a similar form.  As we've discussed in our toy model, a projector of the form given in Eqn.~\ref{eqn:p1def}, which projects only on a system state, creates a product state between system and environment when operating on ${\left| \mathcal I \right\rangle _w}$. The environment state which is correlated in this way is determined by ${\left| \mathcal I \right\rangle _w}$.  Since the state of the environment can contribute to decoherence and other effects on the system (which can impact both the evolution of an individual history as well as interference effects between histories), ${\left| \mathcal I \right\rangle _w}$ plays a more detailed role in the behavior of the histories than in the MH scheme.  Still, once consistent histories are found it would seem natural for observers on a particular history to identify the state at the start of \emph{their} history as the ``state of the Universe,'' rather than ${\left| \mathcal I \right\rangle _w}$.

Bringing this perspective to our ACL model calculations, one can say that the correlation functions calculated in Sect.~\ref{sec:einselect} could just as well have been calculated  starting with the product ``initial'' states which result from operating with $P_0$ on ${\left| \mathcal E \right\rangle _w}$.  Aside from its role in determining which environment state appears in the product, one could simply forget about the role of ${\left| \mathcal E \right\rangle _w}$.  Turning to the CH calculations which followed, looking closely one can see that ${\left| \mathcal E \right\rangle _w}$ plays a role in determining the level of quantum interference among the paths. But if one sticks to the time period where the interference is acceptably low, again the behavior of the path can be described just fine by the product initial state, without direct reference to  ${\left| \mathcal E \right\rangle _w}$.  This disconnect from ${\left| \mathcal E \right\rangle _w}$ offers a helpful context for the fact that we were able to identify plenty of phenomena associated with an arrow of time, despite the equilibrium nature of ${\left| \mathcal E \right\rangle _w}$\footnote{We note that our emphasis on this disconnect is a major difference between this paper and~\cite{Hartle:2020his} (which also discusses the arrow of time, initial conditions and cosmology). }.  This perspective will also prove interesting in our cosmological discussion.

 \section{\label{sec:cosmo}Cosmological Discussion}
 
 \subsection{\label{sec:CosmoBackground}Background}
The goal of this work is to illuminate discussions of cosmological initial conditions. Since the arrow of time figures prominently in such discussions, a result requiring an arrow of time to realize classical behavior would seem to offer important insights. Our calculations have led us to claims that are not quite so simplistic, but as they are we find them all the more intriguing.  

The topic of cosmological initial conditions is a complicated one. There is no universal consensus about what one is trying to accomplish with a theory of cosmic initial conditions, and what features one should require of a successful theory. Some physicists are struck by the apparent tuning that is present in the initial conditions for our observed Universe (which in fact corresponds to the low entropy required to have an arrow of time~\cite{Penrose1979}).  Among those concerned about tuning, some are tempted by the attractor behavior of cosmic inflation~\cite{Albrecht:1985yf,Albrecht:1986pi,Linde,East:2015ggf,Clough:2016ymm,Clough:2017efm} (or of alternative theories~\cite{Cook:2020oaj}) as a tool for dynamically favoring certain initial conditions\footnote{Indeed, students typically emerge from contemporary courses on cosmology with the impression that inflation dynamically resolves all cosmological tuning problems.}. 
Others have argued, based on various phase space considerations, that a dynamical explanation of the early low entropy is impossible~\cite{Eddington1931,FeynmanCharacter,Penrose1989,Goldwirth:1991rj,Gibbons:2006pa} (see~\cite{Brandenberger:2016uzh} for a review of this issue in the context of starting cosmic inflation). And there have been a number of attempts to navigate a more nuanced path among these different points of view~\cite{BOLTZMANN_1895,Albrecht:2002uz,Hawking:2003bf,AlbrechtSorbo:2004ke,Carroll:2004pn,Carroll:2005it,Bousso:2007kq,Albrecht:2014eaa,GuthChamblin2020} (see also discussions at this workshop~\cite{SimonsAtNielsBohr2018}). Yet another school of thought regards the elegance with which one can state the initial conditions more highly than whatever can be accomplished dynamically, for example in certain ``wavefunction of the Universe'' formulations such as~\cite{HartleHawking.PhysRevD.28.2960,Vilenkin:1982de,Brandenberger:2016uzh,Feldbrugge:2017kzv,Brahma:2020cpy,Jonas:2021xkx}. From that standpoint, the low entropy can look like a virtue, rather than a tuning problem.   One could also just take the practical viewpoint that the initial conditions should simply be declared, without fanfare or extensive scrutiny.  This approach might best match how physics is done in fields other than cosmology~\footnote{Quantum gravity, which surely is ultimately the tool we need to address these questions, has a well known ``problem of time'' which has potentially radical consequences~\cite{Albrecht:2007mm}.  As is done in much of the literature on cosmic initial conditions, in this paper we implicitly assume a suitable time variable has been identified (for example along the lines of~\cite{Banks:1984np,Fischler:1985ky}) and pursue an investigation which uses that variable as effectively an external time parameter. We acknowledge that until time in quantum gravity is fully understood it will remain unclear whether our (rather conventional) approach is missing important elements relevant to cosmic initial conditions. }.

This work is motivated by the hopeful view that more thought and technical progress could bring greater clarity and consensus on the topic of cosmic initial conditions.  To connect our ACL calculations with cosmology, we start with some basic comments about the standard big bang cosmology and the arrow of time.  By ``big bang'' we mean a Friedmann-Robertson-Walker (FRW) model adjusted to describe our observed Universe as well as possible. At early times such a model will have small perturbations which form the seeds of galaxies and other cosmic structure that emerges over time due to gravitational collapse around these seeds.  It is currently standard practice to assume such a model emerges after a period of cosmic inflation (or sometimes an alternative dynamical scheme) which accounts for the details of the perturbation spectrum, and perhaps some other aspects, but one could also consider a more ``old school'' picture where the FRW Universe emerges from an initial singularity in the radiation dominated phase with the perturbations simply imprinted from the start. 

In big bang cosmology, the low entropy of the early Universe originates in the FRW form of the metric~\cite{Penrose1979}.  The emergence of cosmic structure (and thus deviations from FRW) via gravitational collapse  is the origin of the arrow of time in the Universe.  As reviewed for example in~\cite{Albrecht:2002uz}, our local instance of cosmic structure (the hot sun radiating into cold space) is the primary origin of the arrow of time we experience here on earth.  Heuristically, it is this instability which prompts concerns about ``fine-tuning.'' Much as one might be surprised to walk into one's office and find a pencil stably balanced on its point, the instabilities of the early Universe reflect an initial balancing act that is at least as striking and mysterious in the eyes of many cosmologists.  While certain classic treatments such as~\cite{Guth:1980zm} focus on the instability associated with curvature within the FRW metric, the more general tuning issue relates to the vast array of other possible metrics that the early Universe apparently ``turned down'' in favor of FRW~\cite{Penrose1979}.

We should note that it is the instability of the early universe to gravitational collapse rather than the FRW metric per se that creates an arrow of time. de Sitter space is also described by an FRW metric, but it is classically stable.  In fact, once notions of horizon entropy are factored in, de Sitter space can be considered the highest entropy state accessible to a universe with a positive cosmological constant~\cite{Gibbons:1977mu}.  In that sense it is a kind of equilibrium state which, as expected for equilibrium conditions, does not exhibit an arrow of time.  The presence of thermal Gibbons-Hawking radiation in de Sitter space~\cite{Gibbons:1977mu} further encourages  an equilibrium interpretation. 

\subsection{\label{CosmACL}Connecting our ACL results to cosmology}
Let us now make some links to our ACL results.  A physics laboratory is an out-of-equilibrium system (ultimately thanks to the arrow of time of the cosmos as a whole). An experimentalist could simply displace a pendulum with their hand and create a situation similar to the one depicted in Fig.~\ref{fig:equillibration}, arrow of time and all.  More sophisticated experiments could measure the correlation functions depicted in Fig.~\ref{fig:ForwardThree}. 
Our experimentalist might also construct a ``Schr\"odinger cat'' superposition of oscillator states and allow interactions with the environment to reflect einselection, as modeled for example in~\cite{ACLintro}.  All of these experiments exploit the cosmic arrow of time, which is available to us in abundance, and illuminate its relationship to the emergence of classical from quantum.  
% (The same arrow of time is also crucial to many other everyday activities~\cite{Albrecht:2002uz}.) 
This paper is motivated by our curiosity about whether the arrow of time is \emph{essential} for the emergence of classical from quantum, particularly with regards to the process of einselection.  Given the extent to which we depend on classical physics in the world around us, it would seem that an answer in the affirmative might provides useful insights about the initial state of the Universe, from which time's arrow originates. 

We should acknowledge here that we have not mapped out a detailed linkage between the arrow of time needed to einselect our SHO and the specific initial state of our observed Universe.  There are many other conceivable initial states which also have an arrow of time to some degree (certainly enough to decohere a single oscillator) but which do not seem as finely tuned.  This point is related to the ``Boltzmann Brain'' problem, which we will return to shortly.  We regard this project merely as a small step in an interesting direction, inspired by these larger questions. 

The direction this step has taken us is something of a surprise.  Rather than disrupting the process of einselection, we have found that using equilibrium states simply drew our attention to the disconnect between the properties of the global initial state and the individual histories experienced by observers.  This disconnect allowed us to consider histories with a clear arrow of time, even though the global state did not exhibit one.  In turn, these out-of-equilibrium histories easily manifested einselection.  The individual histories were far enough removed from the ``detailed balance'' associated with equilibrium that the process of einselection could proceed in the same manner in which it has already been observed in situations which have an arrow of time.  The notion of detailed balance was expressed in the \emph{variety} of histories one could use to interpret the same quantum state, even as many  individual histories had a definite time direction. 

We bring several important basic messages from our ACL studies into cosmology. 
% Most of these are related in one way or another to the fact that a single global quantum state typically can be interpreted using a variety of histories.  
First of all, our work draws attention to the fact that the CH formalism requires one to check for quantum interference effects among histories within a given set, to see which ones can even be assigned classical probabilities.  This point was made long ago~\cite{Gell-Mann:2018dzd, PhysRevD.47.3345}, but it has not been widely implemented. Given the very classical nature of realistic cosmologies, it is unclear to us if this lack of implementation is a serious shortcoming (as argued for example in~\cite{Halliwell:2003fw}).  

Secondly, while quantum physics is able to assign relative probabilities to histories \emph{within} a specific decohering set, it is unable to give a systematic preference to one set over another.  In this sense the different sets of consistent histories represent sets of truly ``alternate facts,'' which describe the same quantum state.  This feature plays an important role in the work presented here, and we reflect further on it in Sect.~\ref{sec:CosmoReflections} and in our conclusions. 

Next, while the global state does have a role in determining the degree of interference among histories, once sufficiently classical histories have been identified the remaining role of the global state is to assign relative probabilities to the different members of the set.  These probabilities have limited meaning to observers who share the same classical history, but they can provide a framework for cosmological discussions of the likelihood of their particular universe\footnote{These features are at least somewhat reminiscent of other work that carefully distinguishes between global and observer perspectives, such as~\cite{Page:2009qe,Srednicki:2010yw,Albrecht:2012zp,Riedel.QMAP.2020}.}.

Finally, we note that our results contradict ideas that ``nothing happens'' in equilibrium states (or even single energy eigenstates as we discussed in Appendix~\ref{sec:EEH}).  Thus we disagree with the application of such ideas to cosmology, as implemented for example in~\cite{Boddy:2015fqa,GuthChamblin2020}.  On this point our arguments seem similar to those which appear in~\cite{Lloyd:2016ahu}. 

\subsection{\label{sec:CosmoReflections}Further reflections}

We have explored the relationship between a global initial state and the perceived initial state experienced by observers on a particular classical history.  Under conditions where interference effects are low, and the history really does look classical, the remaining role of the global state is simply to assign a probability to that history.  One could imagine that cosmologists who come up with a global state which assigns unit probability to the classical path they are on might consider their work finished.  This would correspond to the ``practical approach'' mentioned in Sect.~\ref{sec:CosmoBackground}.  But many other considerations influence people's thinking about a global ``wavefunction of the universe.''  There are cases where cosmologists find these other considerations (essentially ``priors'') compelling enough to favor global states which assign highly suppressed probabilities to the classical paths we are on.  Others are uncomfortable with doing so. This situation reflects the diversity of views about cosmic initial states that we discussed earlier.  To some, willingness to accept a theory in which one's own classical trajectory is assigned a small probability is equivalent to accepting a finely tuned theory\footnote{It's worth noting here that while we've pointed out in Sect.~\ref{sec:AOTother} that the condition of zero bipartite entanglement entropy can be trivially realized on every one of a complete set of paths, at a specific time which can be arbitrarily chosen by suitably choosing the paths, this condition is far from sufficient to provide an arrow of time corresponding to what we see in our observed Universe. Demanding a more realistic condition is likely to highly suppress the associated probability.}.     

To elaborate further, we offer two examples where cosmologists have taken positions in favor of global states where paths exhibiting realistic properties of our observed Universe are exponentially suppressed.   The first is the Hartle-Hawking ``no boundary'' (NB) wavefunction~\cite{HartleHawking.PhysRevD.28.2960}.  It is well known that in theories with an inflaton the NB wavefunction exponentially disfavors cosmologies which experience cosmic inflation in favor of states where the inflaton starts at the bottom of its potential. Nonetheless, proponents of the NB wavefunction find its intrinsic merits\footnote{For example, it has been argued that the NB state dominates the quantum gravity path integral~\cite{Hawking:2003bf}.} sufficient to impose additional conditions which favor inflation to allow more realistic cosmologies to be considered (for example~\cite{Hartle:2015vfa}).  

Another example is de Sitter equilibrium cosmology (dSE)~\cite{AlbrechtSorbo:2004ke,Albrecht:2009vr,Albrecht:2011yg,Albrecht:2014eaa}.   This is a cosmological picture motivated by the idea that the observed cosmic acceleration could be due to a fundamentally stable cosmological constant that defines an equilibrium state for the Universe (along the lines of our discussions of de Sitter space in Sect.\ref{sec:CosmoBackground}).   In that picture, the equilibrium state would be the global quantum state and our observed Universe would be regarded as fluctuation, destined to equilibrate back to de Sitter as we evolve closer to a state dominated by the cosmological constant (conceptually similar to the behavior of the histories we explored with the ACL model).  

Simple arguments suggest that dSE models should suffer from a ``Boltzmann Brain'' problem~\cite{BOLTZMANN_1895,Eddington1931,FeynmanCharacter,Dyson:2002pf,AlbrechtSorbo:2004ke,CarrollBad:2017gkl}. This term refers to the apparent discrepancy between the fact that in equilibrium small fluctuations are much more likely than large ones, yet our Universe appears to be a large fluctuation.  Novel quantum gravity effects could provide a way out of the Boltzmann Brain problem for dSE models~\cite{Albrecht:2014eaa}, but even so fluctuations that resemble our Universe would be Boltzmann suppressed.  An enthusiast of dSE cosmologies might still find a global state dictated by the laws of physics (via equilibration processes) more compelling than one constructed in a more ad hoc manner, and therefore accept the price of Boltzmann suppression.  As discussed in~\cite{Albrecht:2014eaa}, fluctuations like our Universe could be the most likely fluctuations which actually exhibit an arrow of time.  On the other hand, should the exotic phenomena such as proposed in~\cite{Albrecht:2014eaa} not be realized, colleagues who are not willing to favor low probability histories by using theoretical priors may well regard the free availability of out-of-equilibrium histories demonstrated in this work to further enhance the Boltzmann Brain problem.  Such a perspective could extend more broadly to many cosmological scenarios, not just dSE. 

We add one more general thought about dSE models:  While in most of this paper we used the notion of equilibrium as a ``straw man'' to represent the absence of an arrow of time, for dSE models equilibrium is a fundamental part of the physical picture.  If we had concluded that equilibrium conditions prevent the emergence of classicality due to the absence of einselection (as we thought might be the case at the start of this project), that would have created major problems for the dSE picture.  Instead, our results are consistent with identifying histories describing an arrow of time despite the overall equilibrium conditions, as have already been explored heuristically in the dSE literature.

%%%%%%% Changes here for referee response
Another feature of our work that has some connections with cosmology is the presence of a double headed arrow of time. Such ideas come up occasionally in cosmological scenarios (some are discussed in~\cite{Hawking:2003bf,Albrecht:2014eaa,Boyle:2018tzc,GuthChamblin2020}). Here we note some differences between that work and the current discussion.  In our toy model calculations, examples of double headed arrows of time came about by patching together two histories, one defined by an initial condition and the other by a final condition (thus creating a ``middle condition'').  In the context of our CH analysis this patching together makes particular sense in cases where the two paths we are connecting are both behaving very classically at the point of connection, thus extending the classical narrative.  

As emphasized in~\cite{Hawking:2003bf}, the cosmological examples tend not to behave classically at the point where the arrow switches directions.  Instead, the patching tends to occur in a highly quantum regime---often a tunneling event. While there may be reasons to consider wavefunctions that offer a double-headed picture, we note that such a picture is intrinsically different from the cases we have showcased with the ACL model, where the patching occurs at a time of highly classical behavior. The discussion in Sect. 5.2 of~\cite{Albrecht:2009vr} makes the point (which appears to be uncontroversial) that when the ``middle condition'' is intrinsically quantum a discussion of classical phenomena naturally draws the focus to histories with a single arrow, even if technically there is another history with the opposite arrow ``on the other side of'' the quantum domain. 

We conclude these reflections with some general comments about the CH formalism.   We have made extensive use of the feature of this formalism whereby alternate sets of histories (with potentially conflicting narratives) are available simultaneously, providing coexisting alternate interpretations of the same quantum state.  This feature has historically been a source of discomfort and even outright skepticism directed at the CH framework.  One of us (AA) recalls voicing some of that skepticism himself in early discussions of the CH formalism~\cite{Albrecht:1992rs,Albrecht:1992uc}.  In contrast, this paper has fully embraced that feature and it has played a central role in our analysis.  This shift on the part of AA seems partly rooted in a growing appreciation for the limited capacity of quantum physics to answer all questions one might wish to ask (as explored for example in~\cite{Albrecht:2012zp}).  But we also found that our efforts to carefully address the questions posed at the start of this project (particularly as related to detailed balance) drove us to accept and exploit that feature.  

Stepping back a bit, we recognize that the fraught conversations among physicists about the interpretation of quantum mechanics are not about to end. The fact that the nature of the results and motivating questions in this paper nudged one of us into greater acceptance of the CH formalism does not mean others will respond in the same way.  It is certainly reasonable to expect that our results will cause others to become \emph{less} comfortable with that formalism. We have demonstrated histories which seem to give conflicting accounts of the arrow of time (manifesting the well-known capacity for the CH formalism to sustain seemingly conflicting narratives of all sorts). While in this paper we have embraced that feature as a realization of detailed balance, others might regard that feature as evidence that all the histories we consider for equilibrium and stationary systems should be removed from consideration by some enhancement (or outright rejection) of the CH formalism.  The concrete thing we offer is sound technical results which reveal interesting features of the CH formalism and which address topics that are relevant to important questions in cosmology.  We look forward to rich conversations with colleagues who take different viewpoints about the full implications. 

\section{\label{sec:conclude}Conclusions}
We have used a simple toy model to explore the relationship between einselection, the arrow of time and equilibrium.  Einselection, the systematic preference of decoherence processes for special pointer states, is a key element of how classical behavior can emerge in quantum systems. The process of decoherence, or the onset of system-environment entanglement, involves a clear arrow of time in the direction of increased entanglement.  Our work was motivated by the idea that the detailed balance features of an equilibrium state should allow both entanglement increasing and decreasing processes to operate on an equal basis, potentially preventing equilibrium systems from exhibiting einselection. We sought to confirm or deny this idea by investigating whether einselection could exist in an equilibrium system, with the goal of interpreting the implications of such results for the arrow of time in cosmology and cosmological initial conditions.

Furthermore, because the goal of our work was to apply our findings in a cosmological context, we were required  to take extra care to not evoke an external observer in our calculations. Standard correlation function techniques appear to represent measurements by an outside observer, which among other things could reflect a disruption to the assumed equilibrium conditions.  To remedy these concerns we used the consistent histories formalism.  This formalism interprets the evolution of a quantum state in terms of sets of paths, and assigns classical probabilities to paths when quantum interference among the paths is sufficiently low. 

Our calculations reveal interesting relationships among these various ingredients.   We found the consistent histories formalism easily identified paths within the equilibrium system which exhibited an arrow of time, corresponding to a direction of increased entanglement.  Such paths allowed us to explore standard ideas about einselection and show how the physics of entanglement expresses a preference for special pointer states, even within a system that is globally in an equilibrium state.  In contrast to our initial suspicions, detailed balance did not prevent einselection within the equilibrium system in our calculations. Rather, the notion of detailed balance was realized in the diversity of decohering paths with which one could interpret the system—manifesting as disconnected but equally valid sets of paths one could use to describe the identical quantum state. An individual path might express the arrow of time in a particular direction at a given time, but if all the sets of paths were taken together, one would find entangling and disentangling equally represented. We also found sets of paths with double headed arrows of time, and we showed that paths could be found where the point in time at which the arrow switched directions was located at any moment, without preference.  These results led us to carefully scrutinize how limited the influence of global initial conditions is on the physics of individual paths described within the global state, and what this might imply for cosmological initial conditions.

The cosmological context for our work starts with the deep relationship between the arrow of time we experience in the world around us and cosmological initial conditions.  We have reviewed this relationship and also the general challenges faced by attempts to develop a comprehensive theory of cosmological initial conditions.  Placing our results in this cosmological context has yielded a number of interesting insights which we have explored in the previous section.   Specifically, we conclude that one cannot reject cosmological models built on an equilibrium picture based solely on the expectation that classicality is unable to emerge in such theories.  We have explicitly demonstrated counterexamples to such arguments, at least at the level of our toy model. Our results suggest there is no simple way to leverage our practical need for einselection in the world around us to arrive at insights about the global state of the Universe.  The properties of the global state and our experiences on a particular classical history are too disconnected from one another for a simple connection to be made. We've extended our analysis to systems placed in a single global energy eigenstate, and drawn similar conclusions. 

Our work does draw attention to the importance of evaluating the degree of interference among different paths, which if large enough could prevent them from behaving classically.  But our results suggest that the physical features which intrinsically support classical behaviors (such as the weak coupling between the SHO and the environment) also suppress this interference, whether or not the global state has an arrow of time. 

It appears that a broadly agreed upon theory of cosmological initial conditions remains a difficult challenge for the field.
We had hoped our explorations would help this endeavor by exploiting our need for emergent classicality to place limits on how the problem might be approached. Instead, our results draw attention to how disconnected the experiences on one classical history are from the properties of the global quantum state.  While our results do not move things in the direction we expected, being forced to face these implications feels like a certain kind of progress. 

\section{\label{sec:thanks}Acknowledgements}
We thank Patrick Coles and Jonathan Halliwell for helpful conversations. This work was supported in part by the U.S. Department of Energy, Office of Science, Office of High Energy Physics QuantISED program under Contract No. KA2401032.

\appendix

\section{\label{sec:RanPhase}Randomized phases and matrix elements}

The equilibrium state $\left| \mathcal{ E} \right\rangle $  used in our calculations was arrived at (as discussed in Sec.~\ref{sec:eqm}) by evolving an out-of-equilibrium initial state to a time which appears to be deep in an equilibrium regime.  One diagnostic one can try is to take the same state, expand it in eigenstates of the total Hamiltonian ($H_w$)  and completely randomize the phases of the coefficients in this expansion.  If replacing $\left| \mathcal{ E} \right\rangle $ with such a randomized version were to lead to different results, that would signal that artifacts of the out-of-equilibrium initial state could be present in our calculations.  
Another check involves the random numbers generated in the construction of $H_e$ and $H_e^I$.  We should check that our results do not depend on the seed used for the random number generator. 

\begin{figure}[h]
    \centering
    \includegraphics{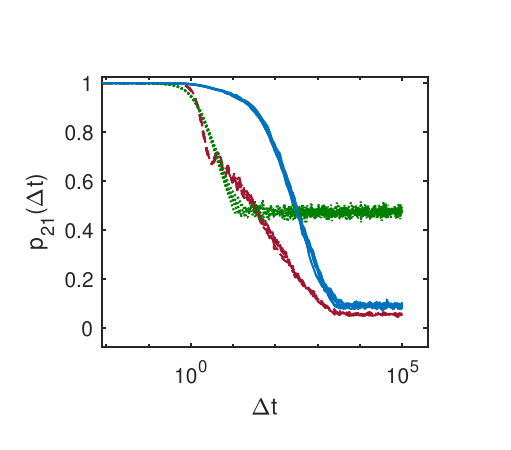}
    \caption{The quantities shown in Fig.~\ref{fig:ForwardThree} are recalculated, each with six variations to the technical details of the calculation. The similarity of each set of six curves with one another (as well as to the curves in Fig.~\ref{fig:ForwardThree}) illustrate the robustness of our definition of equilibrium and a lack of dependence on our random number seed.}
    \label{fig:Randomized}
\end{figure}

Figure~\ref{fig:Randomized} contains six variations on the three curves shown in Fig.~\ref{fig:ForwardThree}. 
Three variations reflect the randomized phase diagnostic, and three used different random number seeds. Each set of six curves differs from one another (and from the curves in Fig.~\ref{fig:ForwardThree}) only in the small-scale noisy features. The broad features on which we based our physical analysis are identical, leading us to conclude that there are no artifacts of the choice of initial state or random number seed in our results.  For simplicity Fig.~\ref{fig:Randomized} only shows some of the quantities examined in this paper, but we have found the other quantities to be similarly well-behaved. 

The equilibration behaviors of the ACL model prompt interesting questions about the relationship between these behaviors and notions of thermalization, the Gibbs distribution, etc. These questions are addressed in~\cite{Albrecht:2022mrx} where it is argued that the qualities exhibited by the equilibration behaviors of the ACL model are sufficient to address the questions posed in this paper (for example by exhibiting detailed balance in equilibrium) even though the toy model nature of the ACL model makes these other notions less generally applicable. 

\section{\label{sec:EEH}Eigenstate Einselection Hypothesis}
The random phase diagnostic discussed above suggests our results reflect quite general properties of the eigenstates of $H_w$, more than specific details of the particular state we chose.  We now follow this path further to see if we can get similar results if we start with a single eigenstate of $H_w$, rather than $\left| \mathcal{ E} \right\rangle $.  This exploration is an extension of the eigenstate thermalization hypothesis (ETH) ideas~\cite{Deutsch_PhysRevA.43.2046,Srednicki_1994} to the topic of einselection.  While the focus of the ETH tends to be multi-particle systems described by field theories, we move ahead here with an exploration in the context of our simple toy model.  The idea that the einselection behavior of a system can be reflected in a single energy eigenstate might be called an ``eigenstate einselection hypothesis'' (EEH).   Regardless of nomenclature, this exploration allows us to challenge claims (such as those in~\cite{Boddy:2015fqa}) that a system in an eigenstate of its total Hamiltonian cannot exhibit interesting dynamics.

\begin{figure}[h!]
    \centering
    \includegraphics{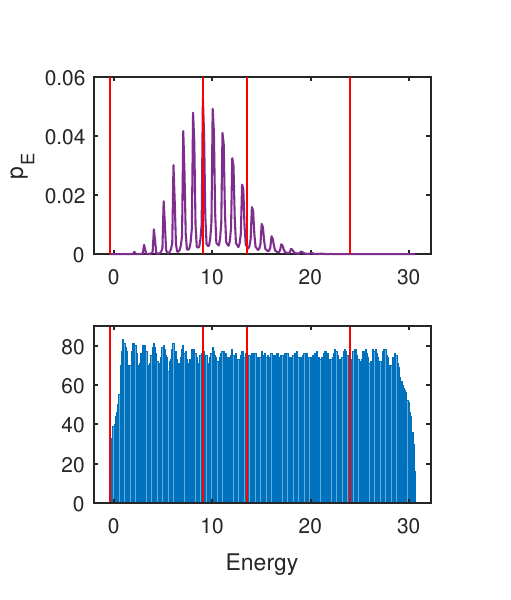}
    \caption{The lower panel gives a histogram of the eigenvalues of $H_w$. The upper panel gives $p_E$, the probabilities assigned to eigenstates in each bin for the equilibrium state $\left| \mathcal{ E} \right\rangle $.  The vertical lines mark the energies of the particular eigenstates of $H_w$ used for Figures~\ref{fig:df245Ea},\ref{fig:df245Eb}, \ref{fig:df245Ed} and \ref{fig:df245Ee}. }
    \label{fig:EWhist}
\end{figure}

As with the ETH, our results depend on which eigenstate of $H_w$ we choose.  To help us navigate among these eigenstates, we start by looking at the spectrum of $H_w$. The lower panel of Fig.~\ref{fig:EWhist} gives a histogram of the eigenstates of $H_w$, and the upper panel shows $p_E$, the probability assigned to each histogram bin in the state $\left| \mathcal{ E} \right\rangle $.
Detailed properties of the spectrum of $H_w$ are discussed in Appendix C of~\cite{ACLintro}.  The oscillatory behavior of $p_E$ reflects the fact that the initial state which equilibrates to generate $\left| \mathcal{ E} \right\rangle $ is a product state with the environment in a single eigenstates of $H_e$.  

\begin{figure*}[htp]
\vspace{-1.4cm}
\centering
\begin{minipage}[b]{0.46\linewidth}
\vspace{0pt}
\includegraphics[width=0.95\textwidth]{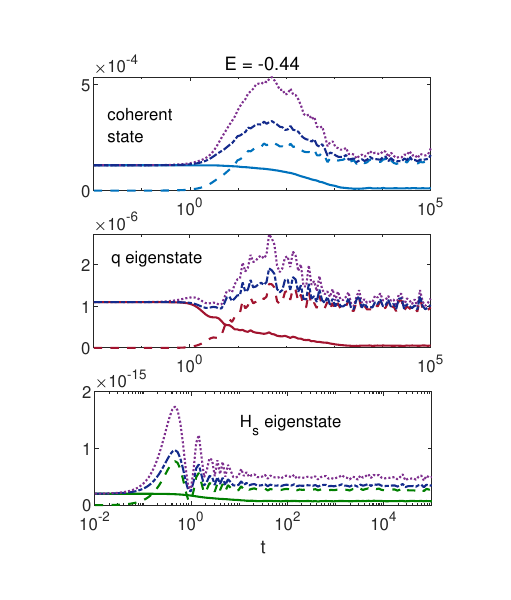}
\vspace{-0.8cm}
\caption{{\small CH quantities as per Fig.~\ref{fig:df245}, but with the equilibrium state replaced with the ground state of $H_w$.  While the solid curve is still most stable in the top panel, giving one signal that coherent states are being einselected, interference among paths (given by the deviation of the dotted and dot-dashed curves) grows sharply at the same time the other panels destabilize.  This suggests that when interference effects are accounted for there is not a strong argument for einselection favoring coherent states in this case.}}
\label{fig:df245Ea}
\end{minipage}
\hfill
\begin{minipage}[b]{0.46\linewidth}
\vspace{0pt}
\centering
\includegraphics[width=0.95\textwidth]{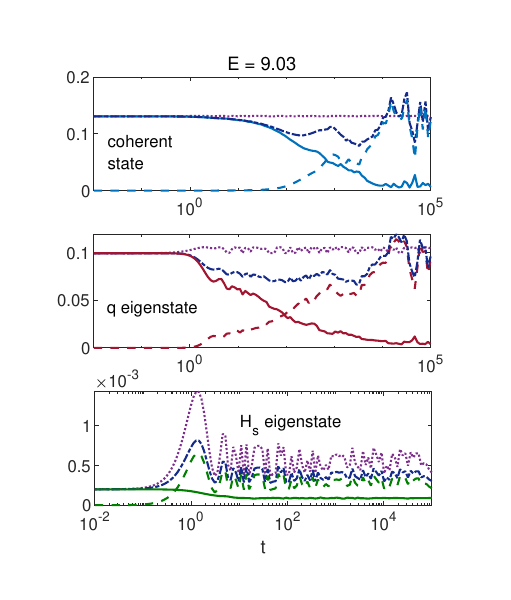}
\vspace{-0.3cm}
\caption{{\small CH quantities as per Fig.~\ref{fig:df245}, but with the equilibrium state replaced with the eigenstate of $H_w$ corresponding to the peak of $p_E$ in Fig.~\ref{fig:EWhist}. This is the eigenstate that has the strongest overlap with the equilibrium state. The quantities in the upper panel are the most stable,  indicating einselection of coherent states is exhibited for this case.} }
\label{fig:df245Eb}
\end{minipage}
\end{figure*}

\begin{figure*}[htp]
    \centering
    \begin{minipage}[b]{0.46\linewidth}
    \vspace{0pt}
    \includegraphics[width=0.95\textwidth]{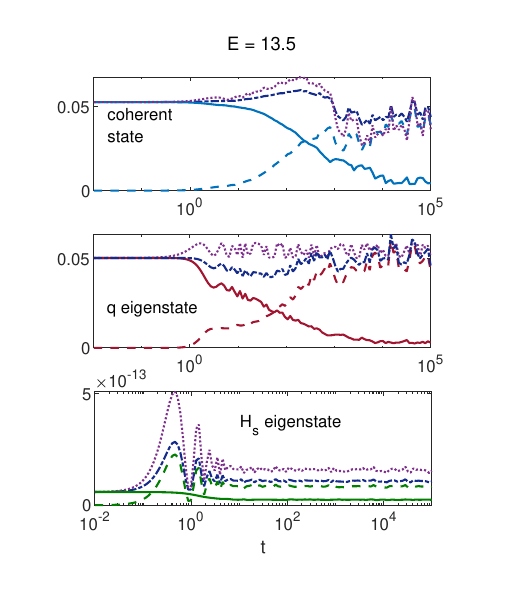}
    \vspace{-0.2cm}
    \caption{ {\small CH quantities as per Fig.~\ref{fig:df245}, but with the equilibrium state replaced with the $E=13.5$ eigenstate of $H_w$. This corresponds to a trough of $p_E$ in Fig.~\ref{fig:EWhist}, but well within the range where $p_E$ is nonzero. The quantities in the upper panel are the most stable, indicating einselection of coherent states is exhibited for this case.} }
    \label{fig:df245Ed}
    \end{minipage}
\hfill
\begin{minipage}[b]{0.46\linewidth}
\vspace{0pt}
    \centering
    \includegraphics[width=0.95\textwidth]{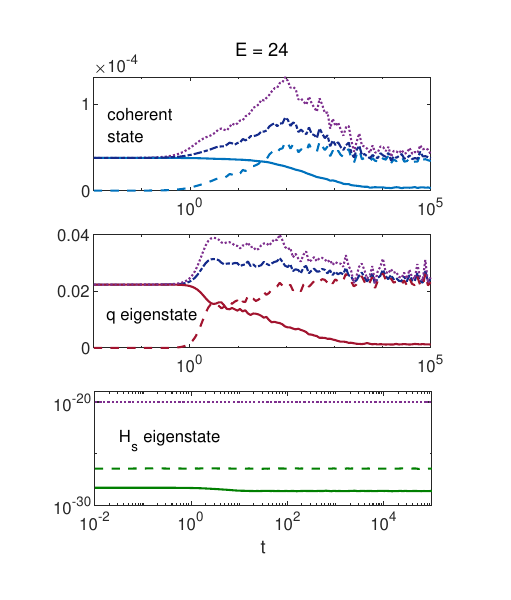}
    \vspace{-0.8cm}
    \caption{ {\small CH quantities as per Fig.~\ref{fig:df245}, but with the equilibrium state replaced with an eigenstate of $H_w$ with $E_w = 24$. This is another state with very little overlap with the equilibrium state. The picture is similar to that in Fig.~\ref{fig:df245Ea}, with some signs of einselection of coherent states shown in the solid curves, but not in the interference effects (given by the deviation of the dotted and dot-dashed curves).  A comment about the anomalous appearance of the lower panel appears in the text. }}
    \label{fig:df245Ee}
    \end{minipage}
\end{figure*}

Figures~\ref{fig:df245Ea},\ref{fig:df245Eb},\ref{fig:df245Ed} and \ref{fig:df245Ee} plot the same quantities shown in Fig.~\ref{fig:df245}, except with $\left| \mathcal{ E} \right\rangle $ replaced with an eigenstate of $H_w$. The corresponding eigenvalue is indicated at the top of each plot, and the locations of these four energy values are marked with vertical lines in Fig.~\ref{fig:EWhist}.  Our broad conclusion based on the four samples shown here as well as additional systematic explorations, is that as long as one chooses an $H_w$ eigenstate which contributes significantly to equilibrium state $\left| \mathcal{ E} \right\rangle $, the general features of the CH quantities are unchanged. In particular, our conclusions about einselection are sustained. In addition, choosing states with eigenvalues from troughs in $p_E$ does not generate significant differences.  This meshes with our explorations of the ACL model, which indicate that starting the environment in a wide range of eigenstates of $H_e$ does not change the overall einselection behavior significantly. The figure captions mention a few additional details. 

Furthermore, we note that Figs.~\ref{fig:df245Ea} and~\ref{fig:df245Ee} both use eigenstates of $H_w$ which have very little overlap with $\left| \mathcal{ E} \right\rangle $. This shows up in the small corresponding values of $p_E$ in Fig.~\ref{fig:EWhist}, as well as the small overall values of the CH quantities (note the small y-axis scales that appear in these two plots). The curves in the lower panel of Fig.~\ref{fig:df245Ee} have especially anomalous behavior. The dot-dashed curve overlaps the dashed curve and is not shown in order to make the figure clearer.  But its location, orders of magnitude away from the dotted curve, signals overwhelming interference effects. We note, as discussed in Appendices B and C of~\cite{ACLintro}, that eigenstates of $H_w$ with large eigenvalues correspond to the larger energy eigenstates of the SHO, which have strange properties due to the truncated nature of the SHO in the ACL model.  But the simplest explanation of the variation in the interference effects across the different cases stems from the different probabilities assigned to the paths.  If the probability assigned to the $10$ path is very small, it does not take much ``leakage'' in from the $1\bcancel{0}$ path to create significant interference effects. Furthermore, when the probabilities assigned after projecting with $P_{\bcancel{0}}$ are larger, there is more overall capacity for such leakage to occur.

\vspace{-0.6cm}

\clearpage

\bibliography{AAlib}% Produces the bibliography via BibTeX.
\end{document}